Chapter 20

# Broken Pair Model: A viable Alterative to the Shell Model for Odd-Spherical Nuclei

I. M *Hamammu, Haq, S., Eldahomi, J. M

*Physics Department, Faculty of Science, Garyounis University, Benghazi, Libya*

## ABSTRACT

The broken pair model has been developed earlier as an useful approximation to the nuclear shell model for even-even nuclei. It is extended and developed here to include odd nuclei too. The model is then applied successfully in the Zr region, using two different sets of matrix elements, the empirical and the realistic sussex ones. The broken pair model results show a considerable agreement with the shell model results wherever available and with the experimental data where shell model calculations are not performed yet.

## INTRODUCTION

The nuclear shell model (SM) has been successfully applied to describe spherical nuclei, but it is well known that as the number of valence nucleons or shells increases the dimensionality of the shell model space increases explosively which make the calculations very difficult, even with today's modern computers (Caurier, 1999, Papenbrock , 2004 , Jie, 2006). Many attempts have been made to overcome this problem by introducing truncation schemes guided by the systematics of the shell model (Schmidt, 1987, Heyde, 1990, Siiskonen, 2000)

The broken pair model (BPM) and its generalized version (Gambhir, Rimini, & Weber, 1969; Rmini & Weber,1970; Gambhir & Haq, 1979; Gambhir, Haq, & Suri., 1981, Allart etal., 1988) is one such method. So far the application of the BPM was restricted to even-even nuclei only. The purpose of this work is to generalize it for odd-even nuclei where both types of nucleons are active in the valence shells. For this purpose, we have derived the relevant expressions for energy levels and transition rates. The model is then applied to calculate the energy levels and E2 transition rates of some nuclei in the Zr region. Two sets of two body matrix elements (m.e.) namely empirical set of Glockner (Glockner, 1975) and renormalized realistic set of Sussex (Ipson, 1978) are used to facilitate a broad comparison with the shell model results. The agreement between BPM and SM results  shows the validity of BPM. The agreement of BPM with the experimental data using both sets of matrix element is encouraging. The departures wherever exist are owed to either excluding an important valence shell or an important higher seniority not included in BPM.

The present work shows that the BPM is a good approximation to the shell model as well as a viable alternative for odd nuclei in the sense that it can be applied when shell model becomes practically too cumbersome.

## 2. The Broken Pair Model:

### 2.1 The formalism:

The broken pair model is based on the assumption that like nucleons; proton-proton (p-p) or neutron-neutron (n-n), prefer to form pairs. The BPM basis states for identical particles in the valence shell are constructed as follows:

The shell model ground state wave function of two identical nucleons is given as [8]

$$|\phi_{00}> = S_+|0> = \sum_a \frac{a}{2} Y_a A^+_{00}|0>, \qquad (1)$$

where $A^+_{00}(aa)$ denotes two particle coupled pair creation operator derived from the two particle coupled creation operator defined by

$$A^+_{JM}(ab) = \sum_{m_\alpha m_\beta} \begin{bmatrix} j_a & j_b & J \\ m_\alpha & m_\beta & M \end{bmatrix} a^+_\alpha a^+_\beta. \qquad (2)$$

Here the square bracket represents the Clebsch-Gordon coefficient and the symbol $a(\alpha)$, $b(\beta)$ represent the single particle states with quantum numbers n,l,j (n,l,j,m). The coefficients $Y_a$ represent the probability amplitude of pair distribution operators over the valence level a. Exclusively they can be defined as

$$Y_a = v_a / u_a \text{ with } u^2 + v^2 = 1, \qquad (3)$$

where $v_a$, $u_a$ stand for the BCS occupation and nonoccupation probabilities respectively.

The approximate BPM ground state wavefunction of p-pairs (2p nucleons) is assumed as (Gambhir, 1978)

$$\tau^+_p|0\rangle = \frac{1}{p} \left[ \prod_a u_a^{j+\frac{1}{2}} \right] S^p_+|0\rangle, \qquad (4)$$

where the parameters $v_a$, $u_a$ can either be obtained by minimizing the Hamiltonian using the above ground state i.e.

$$\delta \left\{ \frac{<0| \tau_p H \tau_p^+ |0>}{<0| \tau_p \tau_p^+ |0>} \right\} = 0, \tag{5}$$

or by solving BCS number and gap equations (Gambhir etal., 1981).

The excited states are then constructed by replacing one pair creation operator $S_+$ in (4) by two particle non pair creation operator defined by (2)

$$\tau_{p-1}^+ A_{JM}^+ |0> . \tag{6}$$

This way by replacing all pair operators, one gets exact shell model Hilbert space.

For odd nuclei, the state of (2p+1) particles are obtained by coupling the odd particle to even particle BPM states. For example

$$\tau_p^+ a_{j_a m_a}^+ |0> \tag{7}$$

shows zero broken pair and

$$\tau_{p-1}^+ \left[ a_{j_a m_a}^+ \otimes A_{JM}^+(rs) \right]_m^j |0> \tag{8}$$

represents one broken pair BPM states of odd nuclei with identical particles in the valence shell.

As a first approximation we assume the BPM basis states for non identical particles occupying valence shells by replacing one of the distributed pair operator $S_+$ of even side of valence nucleons (assumed here neutrons) by an arbitrary two particle non pair creation operator $A^+_{JM}$

$$\left| \phi_{jm}(r_p J_n) \right\rangle = \left[ \tau_p^+ a_{r_p}^+ \otimes \tau_{n-1}^+ A_{J_n M_n}(r_n s_n) \right]_m^j |0\rangle \tag{10}$$

In the next higher approximation the BPM states are obtained by coupling the wave function of one more proton / neutron distributed pair operator $S_+^P/S_+^n$ by one more arbitrary two particle non pair creation operator $A^+_{J_p M_p}/A^+_{J_n M_n}$. In general few particle orthonormal BPM states can be written as

$$\left| \phi_{jm}(d_p j_p d_n J_n) \right\rangle = \left[ X^+_{j_p m_p}(d_p, k_p) \tau_{p-b_p}^+ \otimes X^+_{J_n M_n}(d_n, k_n) \tau_{n-b_n}^+ \right]_m^j |0\rangle \tag{11}$$

where d denotes the additional quantum numbers and k the intermediate angular momenta. $b_p$ and $b_n$ denote the number of broken proton and neutron pairs.

The approximate ground state of odd even nuclei can be written as a product of odd proton / neutron ( p/n pairs + one additional nucleon) (7) and even neutron/proton (n/p pairs) (4)

$$|\phi_{r_p m_p}\rangle = \left[\tau_p^+ a_{r_p}^+ \otimes \tau_n^+\right]_{m_p}^{r_p} |0\rangle \tag{9}$$

Equation (9) has been written with the assumption that protons are odd and neutrons are even (This convention will be followed throughout). However due to proton neutron symmetry the same state and consequently all the expressions can be used for odd neutron and even proton nuclei by interchanging the indices p and n.

## 2.2 The Hamiltonian and its matrix elements:
The Hamiltonian for neutron proton system is given by
H = Hpp + Hnn + Hnp  (12)
where p and n indicates protons and neutrons.

The shell model Hamiltonian for identical nucleons in second quantized form is given as

$$H = \sum_{\alpha_1} \varepsilon_{a_1}^0 \, a_{a_1}^+ a_{a_1} + \tfrac{1}{4} \sum_{\alpha_1 \alpha_2 \alpha_3 \alpha_4} \langle \alpha_1 \alpha_2 | V | \alpha_4 \alpha_4 \rangle a_{a_1}^+ a_{a_2}^+ a_{\alpha_4} a_{\alpha_3} \tag{13}$$

and the np interaction part is given by

$$H_{np} = \sum_{\alpha_p \alpha_p' \alpha_n \alpha_n'} <\alpha_p \alpha_n | V | \alpha_p' \alpha_n' > a_{a_p}^+ a_{a_n}^+ a_{\alpha_n'} a_{\alpha_p'} \tag{14}$$

These equations are to be rearranged in a form suitable for evaluating their matrix elements in terms of overlap of BPM wave functions. Details are given in Ref. [8].

Finally we write (12) as

$$\overline{H} = \overline{H}_{pp} + \overline{H}_{nn} + \overline{H}_{np} . \tag{15}$$

The matrix elements of the Hamiltonian (15) are obtained by

$$\langle \Psi(j_p r_p i_n J_n) | \overline{H} | \Psi(j_p r_p i_n J_n) \rangle = <\overline{H}_{pp}> + <\overline{H}_{nn}> + <\overline{H}_{np}> \tag{16}$$

where $|\Psi(j_p r_p i_n J_n)\rangle$ represents the orthonormalized BPM states for ($b_p$= 0 and $b_n$= 1) defined as

$$\left|\Psi(j_p r_p i_n J_n)\right\rangle = \sum \gamma_{i_p}^{d_p} \gamma_{i_n}^{d_n} \left[a_{r_p}^+ \tau_p^+ \otimes A_{J_n}^+ (r_n s_n) \tau_{n-1}^+\right]_m^j \left|0\right\rangle . \tag{17}$$

Here $\gamma_{i_p}^{d_p}, \gamma_{i_n}^{d_n}$ denote the orthonormalization coefficients for proton and neutron spaces respectively. The final expressions for these terms are given in appendix A.

### 3. Calculations and the Results:

The BPM formalism derived here for odd nuclei has been applied for some nuclei in the

Zr- region. The aim is two fold. To test the validity of the model by comparing its results with the exact shell model results (denoted by ESM) using the same input data and to apply it to some other nuclei in the same region where shell model calculations have not yet been performed.

The ESM calculations have been performed in the past in the Zr-region by Glockner (Glockner, 1975) using empirical set of matrix elements, assuming $^{88}$Sr as an inert core. The model space was restricted to $2p_{1/2}$ and $1g_{9/2}$ proton states and $2d_{5/2}$ and $3s_{1/2}$ neutron states as shown in Fig(1).

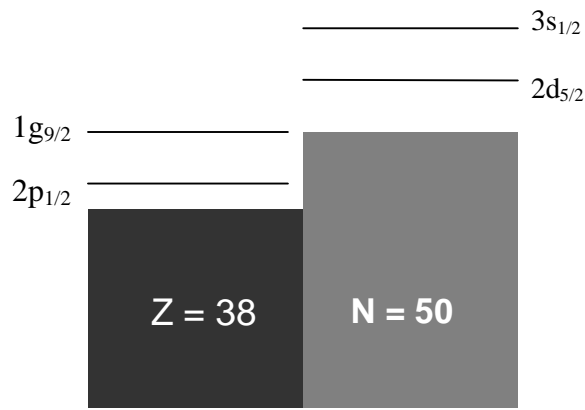

**Fig.1. The proton and neutron spaces used in our calculations.
The shaded area represents $^{88}$Sr core**.

In our calculations we have assumed the same configuration space. First we have used the empirical set of two body matrix elements (denoted by EMP) in order to compare our results with that of the shell model. Next using the same configuration space, we have used the realistic set of matrix elements of Ipson etal (Ipson, Maclean, Booth, & Haigh, 1978) which has been renormalized earlier by Gambhir etal (Gambhir et al., 1981). This set of results is denoted by REAL. Experimental data is denoted by EXPT. The parameters $u_a$ and $v_a$ have been obtained by solving BCS number and gap equations. We present some of the results here. The rest will be published somewhere else.

### 3.1 Identical particle case :

Figs (2) show the results of $^{91}$Nb, $^{93}$Tc and $^{95}$Rh, which correspond to $Z_v=2p+1$ i.e. only one free proton and the rest valence protons were dressed in p_pairs, so that we get only two allowed states just equal to the configuration space. Since our aim is to couple further these proton states with those of neutrons for Y and Nb, it is shown here merely to demonstrate the validity of the BPM for lowest two energy levels as well as to justify our extreme assumption of the first BPM approximation.

Looking at Figs(2), one observes a good agreement between BPM(EMP) and ESM indicating the goodness of the BPM as an approximation to ESM. The quality of the agreement demonstrates the fact that realistic set is as good as EMP in this region. The energy gap between the first two (seniority zero) states and the corresponding higher seniority states for proton active nuclei decreases gradually as we go from Nb to Rh, which means that, the higher seniority components are unimportant, specially for Nb and Rh isotopes as far as low energy is concerned.

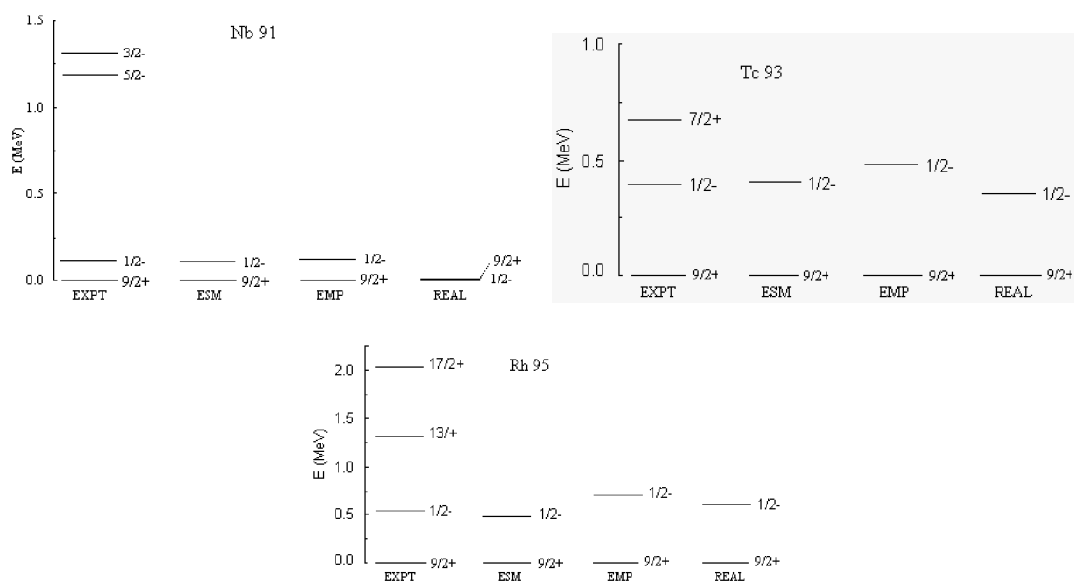

**Fig. 2. Experimental, shell model and BPM energy levels of $^{91}$Nb, $^{93}$Tc and $^{95}$Rh.**

### 3.2 Nonidentical particle case :

### 3.2.1 Odd proton nuclei :

The energy levels of $^{91-95}$Y are given in Figs. (3-5). Fig (3) shows that the results of BPM(EMP) exactly coincide with ESM for $^{91}$Y as it correspond to ( one proton + two neutron states ).

No shell model results are available for $^{93}$Y and $^{95}$Y, so our results will be compared with the experimental data. Figs. (4-5) show how good is the agreement between BPM and the experimental results using both sets of matrix elements. In fact for $^{95}$Y the REAL set is closer to EXPT results than EMP.

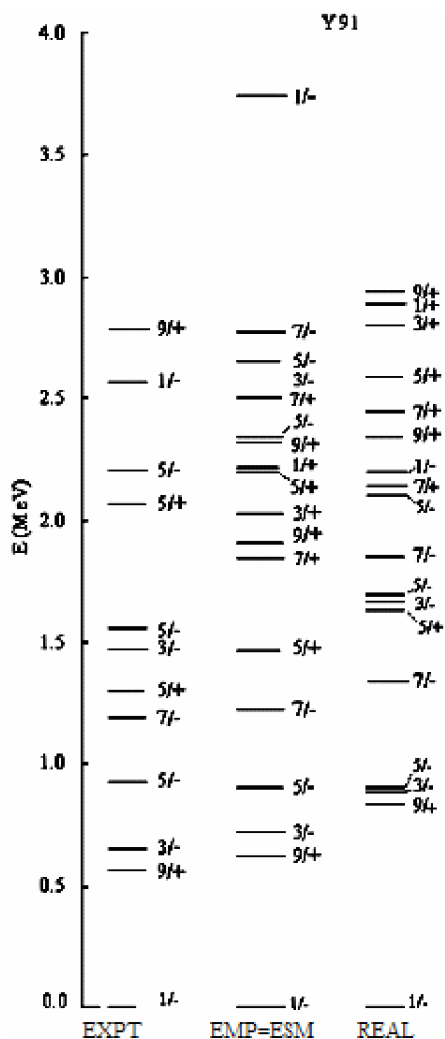

**Fig. 3. Experimental and GBPM energy levels of $^{91}$Y.**

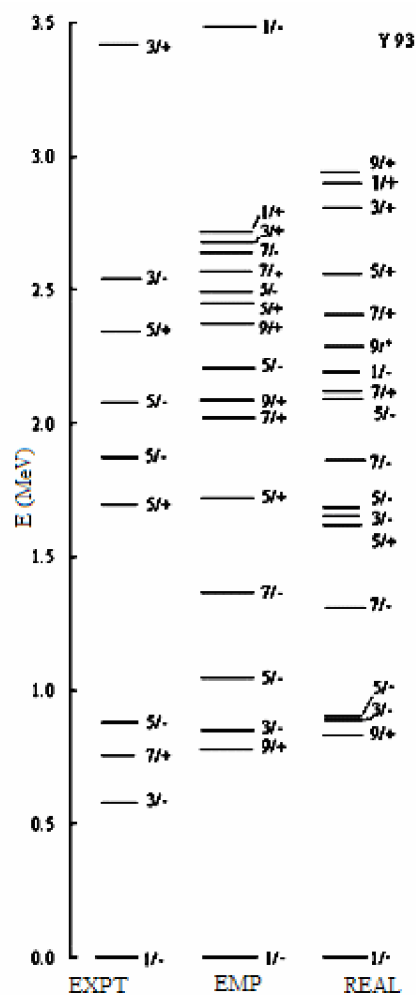

**Fig. 4. Experimental and GBPM energy levels of $^{93}$Y.**

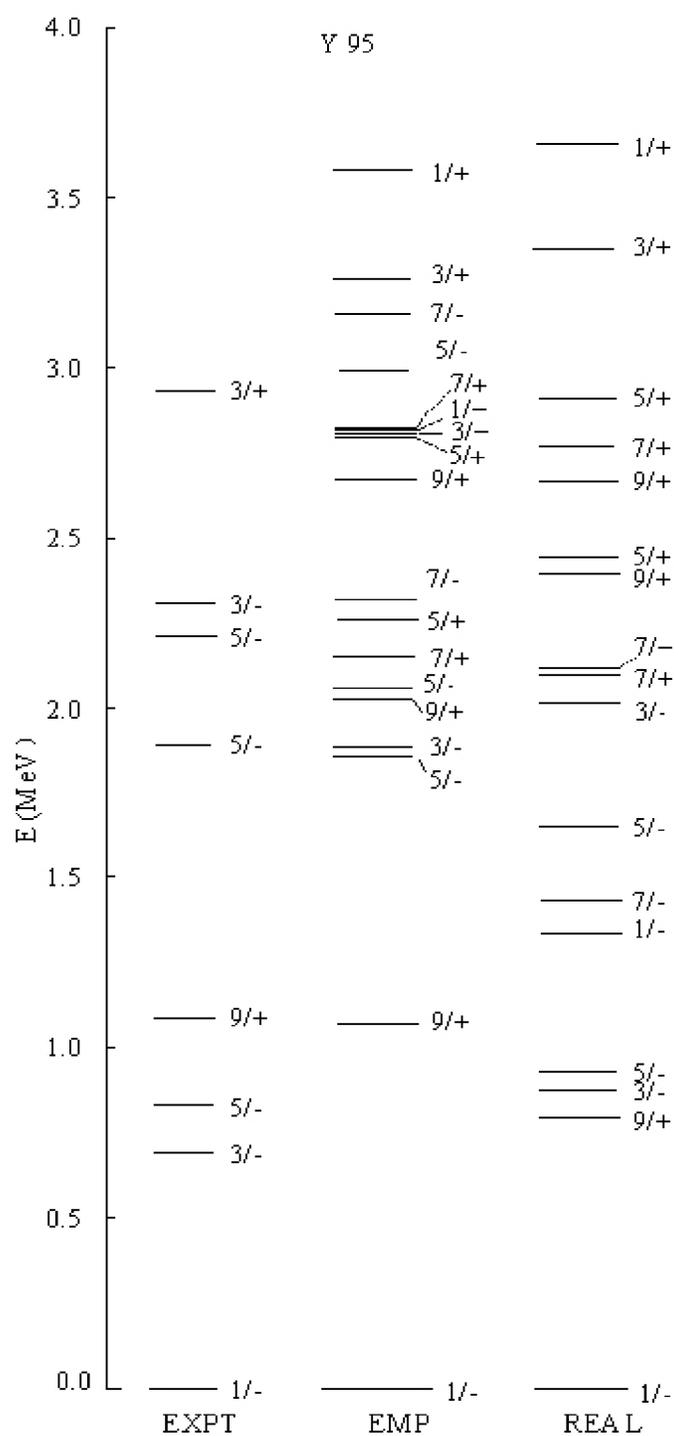

**Fig. 5. Experimental and GBPM energy levels of $^{95}$Y.**

### 3.2.2 Odd neutron nuclei:

ESM results are also available for Zr isotopes. Since there are only two protons and one neutron, in $^{91}$Zr valence levels, hence the results of BPM(EMP) coincide with ESM. It is found from Fig.(6) that the realistic set is in no way inferior to that of EMP, which demonstrates its applicability in this region.

The energy levels of $^{93}$Zr and $^{95}$Zr are shown in Figs. (7-8), from which we note that, in most cases, especially for low energy and high spin states, the BPM(EMP) results compare excellently with those of ESM , except for $3/2^+$ state in which BPM(EMP) compares very poor, with ESM as well as with EXPT. This may be owed to the absence of the interaction of $S_{1/2}$ single neutron state with the neutron occupying $d_{3/2}$ in the valence shells. However in a further calculation including $d_{3/2}$ the BPM results are expected to give good results compared to EXPT. REAL results are again comparable to those of EMP.

It is expected again that the BPM results will improve if $d_{3/2}$ state is also included in the configuration space.

It is to be noted here that, the dimensionality of various nuclear states for the nuclei mentioned here in BPM correspond to three valence shell model states (one p/n + two n/p) for odd proton/neutron nuclei respectively.

## CONCLUSION

The BPM formalism developed here for odd spherical nuclei has been applied successfully in the Zr region. The model results compare fairly well with the shell model results using the same input data. As the results highly depend upon the input data used, two sets of effective two body matrix elements have been used for the sake of broad comparison. The results were found to compare well with the exact shell model results using same input data as well as with experimental results for both sets of matrix elements. This indicates that the BPM can be used when the shell model is practically handicapped due to large dimensionality.

Few discrepancies were found which can be owed to unavoidable higher seniority states. However these discrepancies can be removed in further calculations by breaking one more pair. In some cases some levels will compare well when the model space is enlarged. The results can be improved also by using a new set of matrix elements especially the pairing part of the matrix elements on the odd side.

**Fig. 6. Experimental and GBPM energy levels of $^{91}$Zr.**

**Fig. 8. Experimental, Shell model and GBPM energy levels of $^{95}$Zr.**

# مستويات الطاقة للانوية الفردية-الفردية باستخدام نموذج الازواج المكسورة

**إبراهيم حممو، شمس الحق و جمال الداحومى**

قسم الفيزياء، كلية العلوم، جامعة قاريونس، بنغازي، ليبيا

تقدم هذه الورقة طريقة مبسطة لاستخدام نموذج الازواج المكسورة لحساب مستويات الطاقة للانوية الفردية-الفردية. تم إختبار النتائج بمقارنتها بنتائج نمزذج القشرة فى حالات محددة تطابق فيها نموذج الازواج المكسورة مع نموذج القشرة. بعد ذلك طبق النموذج لحساب مستويات الطاقة لبعض الانوية حول الزركونيوم. تطابقت نتائج النموذج بشكل جيد مع النتائج العملية وكذلك مع نتائج نموذج القشرة فى الحالات التى تم تطبيقه عليها.

مفتاح الكلمات: نموذج القشرة؛ نموذج الازواج المكسورة.